\documentclass[prd]{revtex4}%
\usepackage{amsfonts}
\usepackage{amsmath}
\usepackage{amssymb}
\usepackage{graphicx}%
\providecommand{\U}[1]{\protect\rule{.1in}{.1in}}

\begin{document}
\preprint{HEP/123-qed}
\title[Gravitational wave pulses]{Cylindrical Gravitational Waves in Expanding Universes: Explicit Pulse Solutions}
\author{Robert H. Gowdy}
\email{rgowdy@vcu.edu}
\affiliation{Department of Physics, Virginia Commonwealth University, Richmond, VA 23284-2000}
\keywords{gravitational waves}
\pacs{04.20.Jb,04.30.-w,04.25.Dm,95.30.Sf,98.80.-k}

\begin{abstract}
Solutions analogous to the Weber-Wheeler cylindrical pulse waves are found for
the case of cylindrical gravitational waves in an expanding universe. \ These
pulse solutions mimic the asymptotic properties of waves from an isolated
source in three dimensions and can be used to test the far-field behavior of
numerical simulation techniques.

\end{abstract}
\volumeyear{year}
\volumenumber{number}
\issuenumber{number}
\eid{identifier}
\date{October 3, 2007}
\startpage{1}
\endpage{2}
\maketitle

\section{Introduction}

Exact solutions representing cylindrical gravitational waves in an expanding
(Kasner) empty universe were introduced in an earlier paper.\cite{OpenGowdy}
\ They are potentially useful as models for waves from compact sources in
asymptotically flat spacetimes. One possible application of these exact
solutions is to check how well numerical simulations handle the transition
from the near-field zone to the far-field zone where gravitational radiation
power can be
identified.\cite{CauchyChar1,CauchyChar2,CauchyCharExtract,CauchyCharSphSym}

A shortcoming of the solutions presented earlier is that they are in the form
of integrals over wave amplitudes. Here, we will show that the Bonnor
amplitude distribution\cite{Bonnor} that leads to the Weber-Wheeler
pulse\cite{ERwavesReal} in the Einstein-Rosen cylindrical wave case is
integrable in the expanding-universe case as well. The resulting expression
for a cylindrical pulse in an expanding universe is slightly more complex than
the Weber-Wheeler pulse because it involves an elliptic function of the first
kind, but can still be evaluated with little effort.

The explicit pulse solutions do provide one surprise: The basis functions that
make up these solutions have the desired asymptotic behavior near future null
infinity, but the first solutions found by summing these basis functions do
not. The resulting pulses are not well localized and fill the event horizon of
the expanding universe. \ As a result, they are amplified by the expansion and
fall off as ln$r/r$ instead of at the expected $1/r$ rate. Fortunately,
combinations of these Bonnor-type pulse solutions can produce localized pulse
solutions that are not amplified and have the $1/r$ behavior that is needed to
mimic waves from a compact source.

Section II of this paper provides a brief summary of the key results of the
earlier paper and expands on a brief remark in the earlier paper to display
the Minkowski space background geometry for these solutions. Section III
constructs the basic Bonnor pulse solution and discusses its properties. The
localized pulses that could be used to test numerical algorithms are
constructed and analyzed in Section IV. \ These localized pulses include some
with two peaks, separated by a narrow notch that could be used to test a
numerical simulation's ability to produce accurate waveforms in the far-field region.

\section{Review}

\subsection{The metric}

The general family of exact solutions discussed in the previous
paper\cite{OpenGowdy} have the spacetime metrics%
\begin{equation}
ds^{2}=-e^{2a}dt^{2}+e^{2a}dr^{2}+e^{-2W}r^{2}d\varphi^{2}+e^{2W}t^{2}dz^{2}
\label{metric}%
\end{equation}
where the cylindrical wave function, $W$ is given as an integral over wave
amplitudes and Bessel functions.%
\begin{equation}
W\left(  r,t\right)  =\int_{0}^{\infty}dk\left[  A\left(  k\right)
J_{0}\left(  kt\right)  +C\left(  k\right)  N_{0}\left(  kt\right)  \right]
J_{o}\left(  kr\right)  \label{waveint}%
\end{equation}
The amplitude $A\left(  k\right)  $ corresponds to waves that are regular near
the initial singularity of the spacetime while the amplitude $C\left(
k\right)  $ corresponds to waves that are singular there. So long as the
singular wave amplitudes $C\left(  k\right)  $ satisfy one of the two possible
regularity constraints
\begin{equation}
\int_{0}^{\infty}dk\frac{C\left(  k\right)  }{k}=0;\pi, \label{Abel}%
\end{equation}
the remaining metric function $a$ can be expressed as
\begin{equation}
a\left(  x,t\right)  =-W\left(  x,t\right)  +\int_{0}^{x}\frac{t^{2}Z}%
{t^{2}-r^{2}}rdr, \label{aPlusW}%
\end{equation}
where%
\begin{equation}
Z=\left(  W_{r}^{2}+W_{t}^{2}\right)  -2\frac{r}{t}W_{r}W_{t}-2\frac{r}{t^{2}%
}W_{r}+\frac{2}{t}W_{t}. \label{Zdef}%
\end{equation}
In contrast to the Einstein-Rosen cylindrical wave solutions\cite{ERwavesReal}%
, an integral constraint%
\begin{equation}
\int_{0}^{\infty}\frac{t^{2}Z}{t^{2}-r^{2}}rdr=0. \label{IntegralConstraint}%
\end{equation}
may be imposed to ensure that these hypersurfaces are asymptotically flat
(rather than conical) at spatial infinity.\cite{OpenGowdy}

It is useful, for the purpose of scaling solutions, to assign both $t$ and $r$
the dimension of length. With that assignment, the functions $W$ and $a$ are
dimensionless, $k$ is an inverse length and the amplitude functions $A\left(
k\right)  $ and $C\left(  k\right)  $ scale as lengths.

\subsection{The background spacetime}

The trivial solution $a=W=0$ corresponds to the spacetime metric%
\[
ds^{2}=-dt^{2}+dr^{2}+r^{2}d\varphi^{2}+t^{2}dz^{2},
\]
which may be regarded as describing a background geometry for these solutions.
In terms of the new coordinates, $X,Y,Z,T$ defined by%
\begin{align*}
T  &  =t\cosh z,\qquad Z=t\sinh z\\
X  &  =r\cos\varphi,\qquad Y=r\sin\varphi
\end{align*}
this background metric is just%
\[
ds^{2}=-dT^{2}+dX^{2}+dY^{2}+dZ^{2}%
\]
and corresponds to a sector of flat Minkowski spacetime, as shown in fig. 1.%
\[%
\raisebox{-0cm}{\parbox[b]{8.1121cm}{\begin{center}
\includegraphics[
height=4.2505cm,
width=8.1121cm
]%
{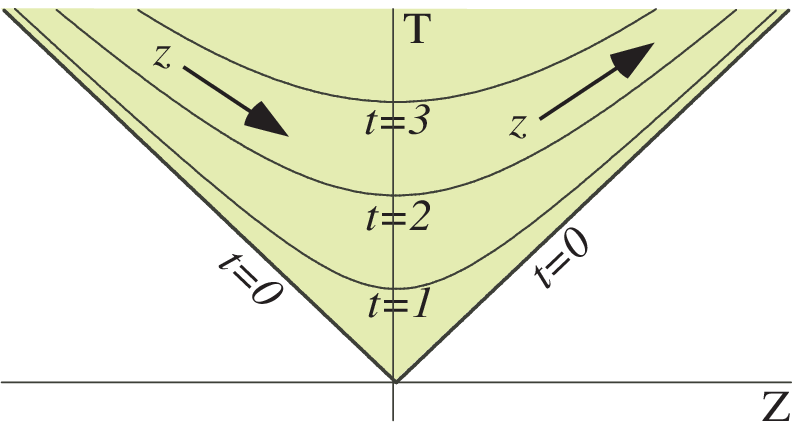}%
\\
FIG.1 The effective background geometry for these solutions is the shaded
region of flat Minkowski spacetime.
\end{center}}}%
\]
In this background geometry, the apparent singularity at $t=0$ is just a
result of using space coordinates that are expanding in the $Z$ direction.
Thus it will sometimes be possible to extend these spacetimes past $t=0$ and
identify translations in $z$ with boosts in Minkowski space. This type of
extension is similar to the Taub-NUT extension and is just the same as the
familiar extension of Misner Space that provides a toy model of the Taub-NUT
extension.\cite{MisnerSpace,MisnerOnNUT,TaubSpace} When such an extension is
possible, the solution will correspond to one of the boost-rotation symmetric
solutions considered by Bi\u{c}\'{a}k and Schmidt.\cite{boostRotation} Even
when an extension past $t=0$ is not possible, these solutions inherit the
future asymptotic structure of Minkowski space, which is one way to understand
how they can be asymptotically flat at future null infinity.\cite{OpenGowdy}

\section{Bonnor Pulses}

\subsection{Integrating the Bonnor amplitude}

The amplitudes $A\left(  k\right)  $ and $C\left(  k\right)  $ are arbitrary,
subject only to the regularity constraint, Eq. \ref{Abel}. The pulse solutions
that we wish to construct correspond to the choices%
\[
A\left(  k\right)  =\alpha e^{-\alpha k}%
\]%
\[
C\left(  k\right)  =0.
\]
where $\alpha$ is the pulse-width (which scales as a length). By choosing
$C=0,$ we solve the regularity constraint trivially and also avoid having to
deal with Bessel functions of the second kind. \ These same choices were
discussed by Bonnor\cite{Bonnor} and later led to Weber-Wheeler waves in the
Einstein-Rosen case.\cite{ERwavesReal} The basic technique for evaluating the
integrals in Eqs. \ref{waveint} and \ref{aPlusW} is also the same as in the
Bonnor-Weber-Wheeler case: \ Use the integral representation of the Bessel
function,%
\[
J_{0}\left(  z\right)  =\frac{1}{2\pi}\int_{-\pi}^{\pi}e^{iz\sin\phi}d\phi,
\]
to turn each integral into a multiple integral and then change the order of integration.

The key result of this paper is the fundamental Bonnor pulse solution
\[
W_{0}\left(  \alpha,r,t\right)  =\alpha\int_{0}^{\infty}e^{-\alpha k}%
J_{0}\left(  kt\right)  J_{o}\left(  kr\right)  dk,
\]
or, using the integral representation,%
\[
W_{0}=\frac{\alpha}{4\pi^{2}}\int_{-\pi}^{\pi}d\zeta\int_{-\pi}^{\pi}d\xi
\int_{0}^{\infty}dke^{k\left(  -\alpha+it\sin\zeta+ir\sin\xi\right)  }%
\]
Two of these integrations are essentially the same as the ones for
Weber-Wheeler waves and leave just one integration still to be performed:%
\[
W_{0}=\frac{1}{2\pi r}\int_{-\pi}^{\pi}\frac{\alpha}{\sqrt{1+w^{2}}}d\zeta
\]
where $w$ is defined by%
\[
w=\frac{\alpha}{r}-i\frac{t}{r}\sin\zeta.
\]
This last integration can be put into the form of an elliptic integral%
\[
W_{0}=\frac{\alpha}{\pi\sqrt{r^{2}+\alpha^{2}}}\int_{-1}^{1}\frac{dy}%
{\sqrt{\left(  1-y\right)  \left(  1+y\right)  \left(  1-\rho^{\ast}y\right)
\left(  1+\rho y\right)  }}%
\]
where%
\[
\rho=\frac{t}{r-i\alpha}.
\]
This elliptic integral can be put into Legendre normal form by the bilinear
transformation%
\[
y=\frac{x+ih}{ihx+1}%
\]
where%
\begin{equation}
h=\frac{1}{2t\alpha}\left(  \left(  r^{2}-t^{2}+\alpha^{2}\right)
-\sqrt{\left(  r^{2}-t^{2}+\alpha^{2}\right)  ^{2}+4t^{2}\alpha^{2}}\right)
\label{soln0}%
\end{equation}
and yields the solution:%
\begin{equation}
W_{0}=BK\left(  m\right)  \label{soln1}%
\end{equation}
where $K$ is the complete elliptic integral of the first kind, the elliptic
modulus $m$ is
\begin{equation}
m=h\frac{h\alpha+t}{ht-\alpha} \label{soln2}%
\end{equation}
and%
\begin{equation}
B\left(  \alpha,r,t\right)  =\frac{2\alpha}{\pi}\sqrt{\frac{h}{t\left(
ht-\alpha\right)  }.} \label{soln3}%
\end{equation}

\subsection{Asymptotic properties and amplification by the expansion of the
universe}

The elliptic function $K\left(  m\right)  $ is just the period of a unit
circular pendulum with $m=\sin^{2}\left(  \theta_{\text{max}}/2\right)  $
where $\theta_{\text{max}}$ is the maximum angle from the bottom of the
pendulum swing. For a large range of values of $m<1$ that period is nearly
constant and almost equal to $\pi/2$. However, the function $K\left(
m\right)  $ has a logarithmic singularity at $m=1$, which corresponds to the
circular pendulum balancing upside-down at the ends of its swing. In the
near-field zone of the solution, where $r$ and $t$ are comparable to the size
parameter $\alpha$, one finds that $m$ is sufficiently smaller than $1$ for
$K\left(  m\right)  $ to be essentially constant. In that region, the function
$B$ describes the pulse quite accurately and the resulting behavior is very
similar to the Weber-Wheeler pulse solution. However, as $r$ and $t$ increase,
the function $m$ rapidly approaches $1$ near $r=t$. The logarithmic
singularity of $K\left(  m\right)  $ then comes to dominate the solution,
which no longer behaves like a Weber-Wheeler pulse.

The peak of a Bonnor pulse occurs at $r=t$ so the peak value can be found by
evaluating $W_{0}\left(  1,r,r\right)  $. Equations (\ref{soln0}),
(\ref{soln1}), (\ref{soln2}), and \ref{soln3} then simplify to \
\[
rW_{0}\left(  1,r,r\right)  =\frac{2}{\pi}\sqrt{\frac{\sqrt{1+4r^{2}}-1}%
{\sqrt{1+4r^{2}}+1}}K\left(  \frac{1}{2r^{2}}\frac{\sqrt{1+4r^{2}}-1}%
{\sqrt{1+4r^{2}}+1}\left(  2r^{2}+1-\sqrt{1+4r^{2}}\right)  \right)  .
\]
An expansion of this expression, using the asymptotic form of the elliptic
function then yields%
\[
rW_{0}\left(  1,r,r\right)  =\frac{1}{\pi}\left(  2\ln4-\ln2+\ln r\right)
+\allowbreak O\left(  r^{-2},r^{-2}\ln r\right)
\]
where the terms proportional to $r$ and $r\ln r$ have cancelled out. The
dominant behavior of the peak is then given by%
\[
W_{0}\left(  1,r,r\right)  =\frac{2}{\pi}\frac{\ln r}{r}+O\left(
r^{-1}\right)
\]

The shapes of these pulses provide an explanation for their anomalous
behavior. \ Figure 2 shows a series of snapshots of a pulse at successive
times. \ This figure plots $tW_{0}\left(  r,t\right)  $ so that an $r=t$ peak
with the expected $1/r$ behavior would appear as a succession of peaks rising
to a constant level. \ The $\ln r/r$ behavior is quite evident in the way that
the peaks grow with $r$. However, even more evident is the fact that these
pulses are not localized and fill the $r=t$ event horizon of the universe.
\ The length scale $\alpha=1$ of the solution characterizes only the radius of
curvature of the sharp peak of the pulse and not the overall width of the
pulse, which is always comparable to the horizon size, $t$. Thus, these pulses
experience a weak amplification because they always have a characteristic
period that matches the time scale of the expansion.
\[%
\raisebox{-0cm}{\parbox[b]{8.3955cm}{\begin{center}
\includegraphics[
height=5.597cm,
width=8.3955cm
]%
{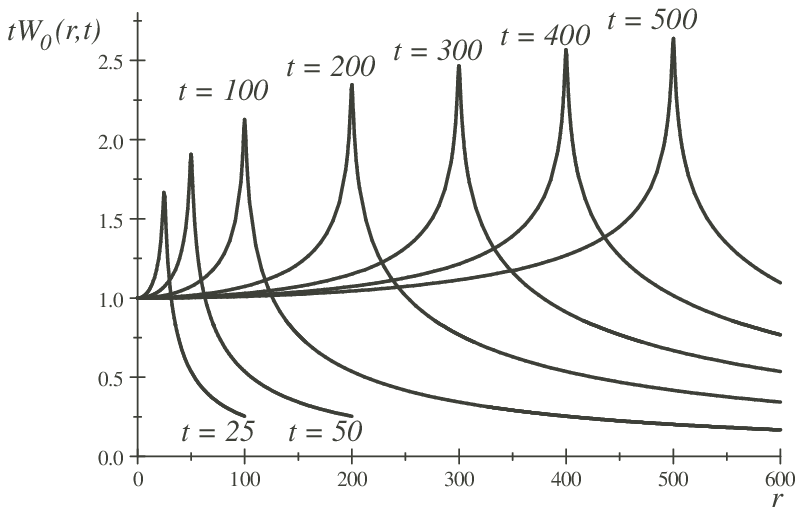}%
\\
FIG. 2 \ Snapshots of a pulse at successive times. \ The pulse grows because
it fills the event horizon and is amplified by the expansion.
\end{center}}}%
\]

\section{Localized, regular pulses}

\subsection{Differences of scaled solutions}

To obtain pulses that are not amplified by the expansion, take advantage of
scale invariance and note that, if $W_{0}\left(  r,t\right)  $ is a solution,
then so are the scaled solutions $W_{0}\left(  br,bt\right)  $ for any
constant $b$. \ The differences%
\[
U_{b}\left(  r,t\right)  =\frac{\pi}{\ln b}\left(  bW_{0}\left(  br,bt\right)
-W_{0}\left(  r,t\right)  \right)
\]
will then also be solutions. These difference solutions are chosen so that the
$\ln r/r$ terms cancel, leaving only a $1/r$ behavior for large values of $r$.
The normalization is chosen so that the peak value will approach exactly
$1/r.$ Figure 3 shows a series of snapshots of the difference solution for
$b=2$. The $r=t$ peak in $tU_{2}\left(  r,t\right)  $ now approaches a
constant value of $1$, indicating a $1/r$ dependence that is no surprise since
we built that property in to these solutions. As a confirmation that we
understand how these pulses work, the figure also shows that the peak is
clearly localized near $r=t$ and has a width that is comparable to the length
scale $\alpha$ (taken to be $1$ in the figure). \ For these pulses, the
characteristic time is a constant $\alpha/c$ so that there is no continuing
resonance with the expansion to amplify them. \ By choosing peaks that are not
amplified, we have obtained peaks that are also localized.%

\[%
\raisebox{-0cm}{\parbox[b]{8.4285cm}{\begin{center}
\includegraphics[
height=4.9907cm,
width=8.4285cm
]%
{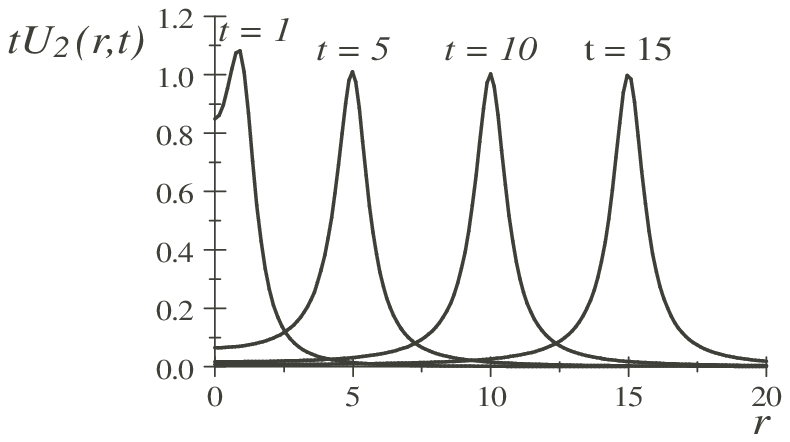}%
\\
FIG. 3 \ Snapshots of a difference pulse at successive times. \ This type of
pulse is localized and falls off as 1/r.
\end{center}}}%
\]
Another way to view the behavior of these difference solutions is to plot the
peak values or $rW$ versus $r$ \ for different values of the scaling
parameter, as in Fig. 4. \ There it can be seen that all of these solutions
begin a $1/r$ behavior within a distance $\alpha$ (taken \ to be $1$ in the
figure).
\[%
\raisebox{-0cm}{\parbox[b]{8.3955cm}{\begin{center}
\includegraphics[
height=5.5948cm,
width=8.3955cm
]%
{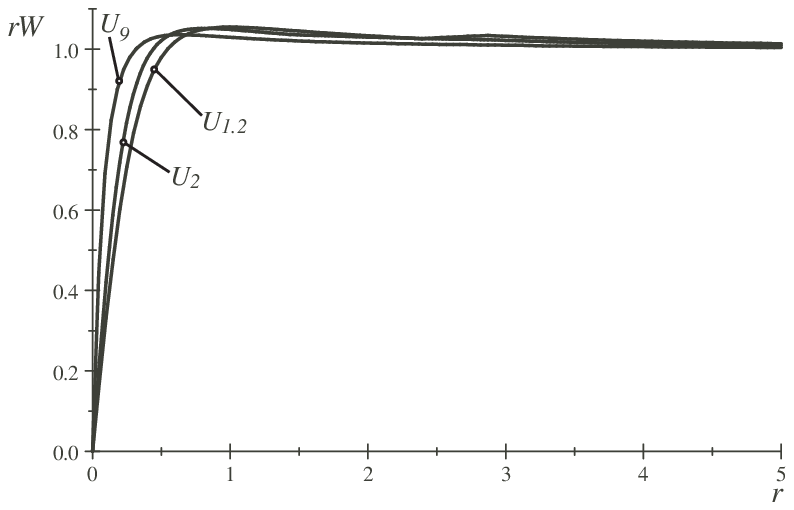}%
\\
FIG. 4 \ Peak values of difference solutions for three values of the scaling
parameter.
\end{center}}}%
\]

\subsection{Notched waves}

Each of the difference solutions consists of a single pulse. \ More complex
solutions with multiple pulses can be constructed by combining difference
solutions with different scaling factors. \ Figure 5 shows a snapshot of a
solution that consists of the difference%
\[
W=U_{2}-U_{9}%
\]%
\[%
\raisebox{-0cm}{\parbox[b]{8.3955cm}{\begin{center}
\includegraphics[
height=7.396cm,
width=8.3955cm
]%
{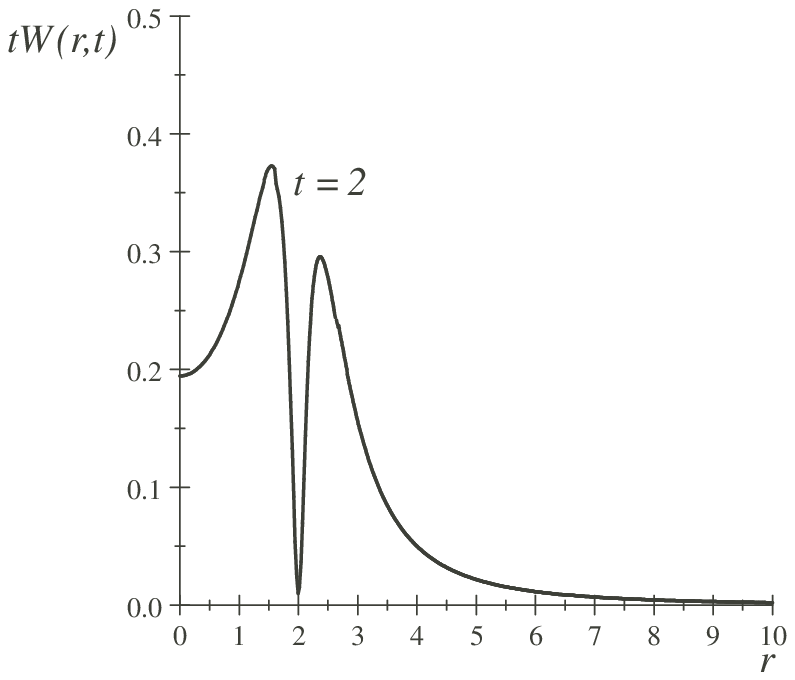}%
\\
FIG. 5 \ A \textquotedblleft notched wave\textquotedblright\ formed by the
difference of scaled solutions $U_2-U_9$.
\end{center}}}%
\]%
\[%
\raisebox{-0cm}{\parbox[b]{8.3955cm}{\begin{center}
\includegraphics[
height=7.3982cm,
width=8.3955cm
]%
{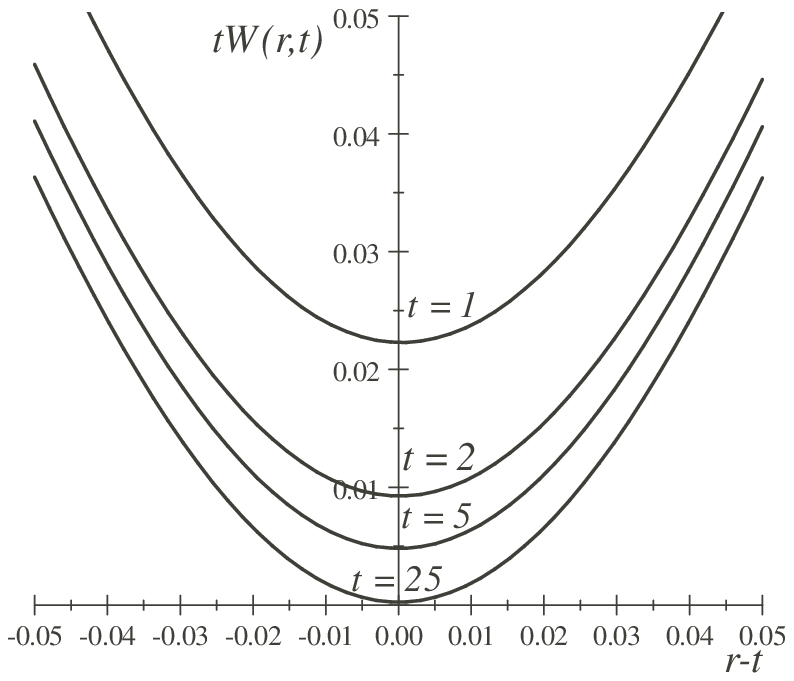}%
\\
FIG. 6. \ Magnified view of the bottom of the "notch" in the $U_2-U_9$
solution at four values of the time, $t$.
\end{center}}}%
\]
The most striking feature of this solution is a deep and narrow trough or
\textquotedblleft notch\textquotedblright\ between two maxima of the
gravitational wave amplitude $W$. As the magnified view in Fig. 6 shows, the
bottom of this notch is actually smooth but the second derivative of $W$, and
thus the spacetime curvature, is extremely large there. As Fig. 6 also shows,
the bottom of the notch moves downward to the axis at large values of $t$.
This type of exact solution could be used as a sensitive test of a numerical
simulation's ability to reproduce exact waveforms in the far-field region.by
observing the behavior of the notch there.

\section{Discussion}

The localized, regular solutions described here represent short pulses of
gravitational radiation that spread in a way that mimics radiation from a
compact source in three dimensions. \ The waves are cylindrical, which
accounts for spreading in two dimensions. The waves inhabit a universe that is
expanding along the cylindrical axis, which accounts for spreading in the
remaining dimension. \ To use one of these solutions to test the far-field
behavior of a numerical simulation of spacetime dynamics, one would evaluate
the solution at a time $t<\alpha$ to find initial values of the metric
functions and their time derivatives as functions of $r$ and would then use
the numerical simulation scheme to evolve the spacetime to $t>>\alpha$. In the
simplest case, where no explicit gravitational wave extraction scheme is used,
the wave pulse would just run off the edge of the numerical evaluation grid,
where one would typically have some sort of damping scheme to minimize
spurious reflection. A comparison to the exact solution would then make it
possible to see just how much spurious reflection actually occurs. \ More
complex schemes such as Cauchy Characteristic
Matching\cite{CauchyChar1,CauchyChar2,CauchyCharExtract,CauchyCharSphSym}
could be tested in a similar way for unwanted reflections from their matching
regions. The notched solutions should be particularly useful for finding
accumulated numerical errors in the far-field waveform produced by any of
these simulation schemes.

\bibliographystyle{apsrev}
\bibliography{acompat,gowdy}

\end{document}